\documentclass[aps,prd,showpacs,twocolumn,floatfix]{revtex4}
\usepackage{graphicx}
\usepackage{amsfonts}
\usepackage{amssymb}
\usepackage{amsmath}
\usepackage{flafter}
\usepackage{epstopdf}
\usepackage{hyperref}
\usepackage{xcolor}
\usepackage{float}
\usepackage[stable]{footmisc}

\begin{document}

\title{Inflation as a Solution to the Early Universe Entropy Problem}

\date{\today} 
\author{Pisin Chen$^{1,3,4}$}
\email{pisinchen@phys.ntu.edu.tw}

\author{Po-Shen Hsin$^{2,3}$}
\email{r01222031@ntu.edu.tw}

\author{Yuezhen Niu$^5$}
\email{yuezhenniu@gmail.com}

\affiliation{1. Department of Physics and Graduate Institute of Astrophysics, National Taiwan University, Taipei, Taiwan 10617}
\affiliation{2. Department of Physics, National Taiwan University, Taipei, Taiwan 10617}
\affiliation{3. Leung Center for Cosmology and Particle Astrophysics (LeCosPA), National Taiwan University, Taipei, Taiwan, 10617}
\affiliation{4. Kavli Institute for Particle Astrophysics and Cosmology, SLAC National Accelerator Laboratory, Menlo Park, CA 94025, U.S.A.}
\affiliation{5. Department of Physics, Massachusetts Institute of Technology, Cambridge, MA 02139, U.S.A.}
\addtocounter{footnote}{-20}

\begin{abstract}

There exists the `entropy problem' of the early universe, that is, why did the universe begin with an extremely low entropy and how did it evolve into such high entropy at late times? It has been long believed that inflation cannot be the solution since it requires an extremely low entropy to ever occur. However, we point out that since the inflation is always accompanied with a horizon, the correct probability of inflation is associated with the quantum entanglement entropy, which should in principle be larger than what considered previously. This motivates us to reexamine the issue by computing the evolution of the cosmological entanglement entropy in the early universe. We invoke a toy model of nonlinear generalized Chaplygin gas (GCG), which has the advantage of providing a smooth and unitary transition between the inflation epoch and the radiation dominant era. We found that soon after the onset of the inflation, the total entanglement entropy rapidly decreases to a minimum, and it rises monotonically afterwards throughout the remainder of the inflation and the radiation epochs. This indicates that the universe does not need to begin with an extremely low entropy; its smallness can be naturally induced by the dynamics of inflation itself. We believe that our computation largely captures the essential feature of entropy evolution and can provide us insights beyond the toy model. 

\vspace{3mm}


\end{abstract}

\maketitle

\section{Introduction}

An increasing amount of observational evidence supports the notion that the early universe has undergone an epoch of inflation. However we are still far from understanding the underlying assumptions and resolving some of its most crucial issues. One of the open questions about inflation is the entropy problem\cite{Penrose1979}\cite{Wald2006}\cite{Sean2010}\cite{Andreas}: why did the universe start from an extremely low entropy initial state that in turn can later give rise to the arrow of time in accordance with the second law of thermodynamics?

Thermodynamic arrow tells us that a system will evolve towards the increase of entropy, which simply states that a system is more likely to evolve from a state with low occurrence probability to that with a higher probability. Since the universe is old, in order to explain the thermodynamic arrow of time the entropy in the early universe must therefore be extremely small, which requires a special initial condition.

On the other hand, the standard inflation paradigm assumes the initial condition of the universe to be realized generically. This conflict \cite{Andreas} becomes more acute if we regard the initial entropy as the probability for spontaneous formation of a homogeneous domain in the inflationary universe. An extremely low initial entropy required by the thermodynamic arrow then suggests that the initial universe only has a very small probability to evolve into the current universe through inflation. 

Several solutions have been suggested in the literature. One approach is the spontaneous eternal inflation \cite{Sean2010}\cite{Sean2004}, where inflation is assumed to occur both forward and backward in time. Another is the so-called bubble cosmology\cite{Coleman1980}, where universes were induced as a realization of string landscape\cite{McInnes2007}\cite{McInnes2008}\cite{Bousso}. 

In this paper, we seek a solution to the entropy problem within the present framework of the standard model of cosmology without relying on assumptions like symmetric time arrows or string theory. The framework will only be effective, in that new physics is expected to occur near the initial singularity predicted by General Relativity, and detailed mechanism for the transition from inflation to a radiation-dominated universe is still not well-understood.

Our aim is to establish the following three stages of entropy development in the very beginning of the universe:
\begin{enumerate}
\item Inflation occurred with a probability that is higher than what is commonly believed.
\item Subsequently a lower entropy state is dynamically induced by inflation itself.
\item A large amount of entropy is generated after it reaches a minimum during inflation.
\end{enumerate}
Note that an important subtlety related to the second stage is that the drastic expansion at the very beginning of the universe is a highly non-equilibrium process. Therefore the usual thermodynamical laws do not necessary hold. The thermodynamic arrow of time is meaningful only after some time span when we come to the third stage, only by then do we have the entropy problem to concern with.

The possibility of this scenario has long been refuted at stage 1 in the previous literature \cite{DonPage}\cite{HawkingPage}\cite{CarrolChen}\cite{Greene} based on the argument that inflation itself requires an extremely special initial condition to occur. However, we hold the point of view that whether the initial condition is special or not depends largely on how one measures the probability of the degrees of freedom. For example, in the recent development of loop quantum cosmology \cite{AshtekarSloan} it was found rather differently that the probability for inflation to happen is much higher using their measure. We should like to point out an important fact, which was overlooked by those previous authors, that {\it inflation is accompanied with a horizon}
\footnote{
Notice that here we directly measure the probability of {\it metric configuration} with inflationary character, instead of imposing condition on source to trigger inflation, therefore there is no danger about ``mismatch in time''.
}.
Therefore the correct measure of the probability for inflation should be given by quantum {\it entanglement entropy}. Since the presence of the horizon increases the entanglement entropy by the entanglement between the observable and the unobservable regions, inflation should have a higher probability to occur. This insight motivates us to seek new dynamical solution to the entropy problem within the inflation paradigm.

We notice that in the literature there are examples of decreasing entanglement entropy \cite{ASeanJohnson2011}\cite{Tegmark2011} due to the situation where the system is out of equilibrium or simply from long-range quantum correlation. In particular, a decrease of tripartite entropy after inflation is shown in \cite{Tegmark2011}, which inspires us to consider the evolution of bipartite entropy from quantum entanglement. Indeed, in a recent effort by two of the present authors (PC \& YN) \cite{Y}, it was found that a minimum of the entanglement entropy induced from the entanglement between cosmological perturbations is reached at an early stage of the inflation. The calculation made there, however, was classical. Since we expect quantum effects to be prominent in the beginning of the universe, each degree of freedom must subject to quantum fluctuations. We therefore re-address the issue properly in this paper.

This paper is organized as follows. In section \ref{entropy} we briefly summarize the reasons and methods for considering entanglement entropy as entropy budget of early universe, which consist of homogeneous and inhomogeneous parts. In section \ref{chap} we introduce the effective model used in our calculation. Detailed treatment and numerical results are summarized in next two sections \ref{WDW} and \ref{pertSect}, which together consist of the total contribution to entanglement entropy. To make comparison with the previous result in \cite{Y}, the computation therein is reproduced in section \ref{pertSect} for completeness. Results in each section are discussed, and possible physical origins of the entropy decrease are presented in section \ref{sec:explain}. The main conclusion is summarized in the last section.

\section{Entanglement Entropy as Entropy Budget}\label{entropy}

\subsection{Why Entanglement Entropy?}

In the present framework, most conventional considerations of entropy are thermal, statistical and entanglement entropy accounted by von Neumann. Since early universe experienced accelerated expansion, it is far out of equilibrium, and thermal entropy is ill-defined for lack of a universal temperature. Statistical mechanics is also not directly applicable on present situation due to long-range nature of gravitational force. 

On the other hand, von Neumann entropy from quantum entanglement is perfectly well-defined in non-equilibrium system, and can cope with problems statistical entropy encounters in gravitational systems like loss of information. Moreover, von Neumann entropy fits into unitary framework of entire universe in that it measures information encoded in entanglement between subsystems, while thermal entropy is of dissipation nature.
Since the early universe can be described as a quantum system without strong decoherence, it is justifiable to invoke von Neumann entropy to find the evolution of total entropy of the universe.  We will thus use the quantum entanglement entropy instead of thermal entropy in the following discussion to help clarify our setup and to quantify the amount of information in a quantum system.

Here we will clarify what we mean by von Neumann entanglement entropy. Entanglement entropy of von Neumann measures the degree of bipartite entanglement, and thus depends on partitions of the system.
Multipartite entanglement entropy thus seems necessary in order to characterize total entanglement entropy of a system; however, the method is difficult to applied on general systems \cite{Wei2003}.
We suggest one reasonable way to quantify the total entanglement entropy of a quantum system is to sum up von Neumann entanglement entropies of all bi-partite subsystems. Since we only concern the entropy change in the evolution, we can subtract a ground value to regularize any potential infinity. More on regularization will be discussed in later sections.

\subsection{Entanglement Entropy Budget}

In the construction of entanglement entropy evolution, there are two contributions arose from two distinctive ways of dividing the universe. Regarding the entire universe as unitary quantum mechanical system consist of observable and unobservable regions, von Neumann entropy then measures the quantum entanglement between those two parts of universe due to the evolution of background metric, and is thus a homogeneous contribution. In this unitary setting von Neumann entropy characterizes the probability of different histories, since the total amount of information is kept constant by Louville's theorem. The entanglement entropy is then obtained by the standard Wheeler-DeWitt formalism, where configurations of fields represent different histories in the path integral sense. We will discuss homogeneous entropy from particle horizon in more detail in section \ref{entrophori}.

The second contribution comes from the entanglement between two opposite momentum sectors of the universe, and is entitled inhomogeneous entanglement entropy. In our model of homogeneous background, it attributes to entanglement of pairwise matter momentum modes, which can be computed by standard cosmological perturbation. The total inhomogeneous entanglement entropy is then obtained by summing over the entire momentum spectrum.

\subsection{Entanglement Entropy from Particle Horizon and Area Law}\label{entrophori}

In this section our discussion will use FRW metric and minisuperspace. The Hilbert space in the accessible and the inaccessible regions are distinguished by the scale factor $a>0$. 
The concept of particle horizon can be most easily illustrated by considering a light path:
\begin{equation}
ds^2=0=dt^2-a(t)^2d\mathbf{x}^2.
\end{equation}
Thus at infinite future the light can travel a coordinate distance
\begin{equation}\label{eqn:lightdistance}
r_H[a]=\int_{t_0}^{\infty} \frac{\mathrm{d}t}{a(t)}.
\end{equation}
It is therefore clear that if the integral Eq.(\ref{eqn:lightdistance}) converges, light can only travel a finite distance, and we have a particle horizon outside this range; Inflation is one such case.


The entanglement entropy of the universe can be obtained by tracing out the Hilbert space associated with the inaccessible region beyond the particle horizon, i.e., the region with radius $r>r_H$, with $r_H$ given by Eq.(\ref{eqn:lightdistance}). Every such radius gives rise to a concentric sphere in this region, which can always be associated with a different scale factor $\tilde{a}(t)$, evolved from $t_0$, that gives a larger integral than $r_H[a]$. To illustrate this, let us consider different time slices along the curves of $a(t)$ and $\tilde{a}(t)$. At each time slice we can choose $\tilde{a}$ smaller than $a$. Thus the integral in Eq.(\ref{eqn:lightdistance}) for $\tilde{a}(t)$ will be larger than that for $a(t)$, therefore light in this metric can travel beyond $r_H[a]$ under the original metric $ds^2[a]$. We can choose the class $\{\tilde{a}\}$ to be in one-to-one correspondence with each coordinate distance $r$ beyond $r_H[a]$. Thus the trace over the Hilbert space associated with the inaccessible region is performed over those configurations that belong to this class:
\begin{equation}
\mathcal{R}=\left\{\tilde{a}\;|\;\tilde{a}(t)<a(t)\;\forall t\right\}\cong \left\{\tilde{a}\;|\;r_H[\tilde{a}]>r_H[a]\right\}.
\end{equation}
In the language of minisuperspace quantum mechaincs, the trace is then carried out by summing over states with metric degree of freedom $\tilde{a}<a$. The detailed calculation is implemented in section \ref{WDW}.

Before ending the discussion on entanglement entropy, we should mention that the entanglement entropy of the universe has long been considered in the literature, see e.g. \cite{Srednicki}\cite{Muller}\cite{EntdS}\cite{GDSMC}\cite{2012}\cite{Hologbraneworld}\cite{Profent}\cite{Holoent}\cite{Holographent}\cite{Susskind}\cite{Chen}.
For a relativistic quantum field theory in $(3+1)$-dimensional spacetime, the entanglement entropy obtained by partitioning the system into $A$ and $B$ with a smooth and compact boundary $\partial A$ exhibits the following scaling character:\cite{Holographent}
\begin{align}\label{eqn:ententropyscaling}
&S_A=c_1(l/b)^2+c_2\ln(l/b),
\end{align}
where $l$ is the typical length of $\partial A$, $b\to 0$ is some cutoff lattice length, and $c_1>0,c_2<0$ are some constants \cite{Holoent}\cite{Susskind}\cite{Chen}. This relation has been confirmed using Weyl anomaly, and the same expression also appears in the correction obtained by generalized uncertainty principle\cite{Chen}. Note that the leading divergence of the entanglement entropy from (\ref{eqn:ententropyscaling}) is proportional to the area of $\partial A$, which is the ``area law'' of entanglement entropy. In particular, in the presence of horizon the bi-partition can be naturally chosen as inside and outside the horizon, therefore the entanglement entropy should be proportional to the horizon area, which can be regarded as the Bekenstein area law (a more correct statement is that the entanglement entropy should be thought as quantum correction to Bekenstein-Hawking entropy due to matter field, see \cite{Susskind}). 

In our setting the scale $l$ is the cosmological scale at the horizon $l=a$. Let's consider the change of sub-leading term with respect to this cosmological scale:
\begin{equation}
\frac{\mathrm{d}}{\mathrm{d}a}c_2\ln(a/b)=\frac{c_2}{a}<0.
\end{equation}
Therefore if the sub-leading term is non-negligible, it can generate a decrease in entanglement entropy with respect to the cosmological time $a$.

In section \ref{arealaw} we will show the entanglement entropy calculated by our method agrees with the area law in the semi-classical regeme; however, at early times the area law will be corrected by quantum fluctuations (section \ref{qearlyentropy}). Thus our result suggests an {\it ab initio} confirmation of equation (\ref{eqn:ententropyscaling}) from canonical quantum gravity.

\subsection{Choice of Model}\label{choice}

To track entropy evolution, we find it necessary to specify our setting, so long as it is general enough to characterize the essential features of the evolution of the early universe. The early universe we are investigating at can be roughly divided into two stages: inflationary and radiation dominated eras. From the requirement of unitarity, our model must provide smooth transition between those stages. In addition, we will work in the conventional FRW homogeneous background. An candidate with minimal ad hoc assumptions for our mathematical model is generalized Chaplygin gas (GCG), which can provide a smooth transition via its equation of state. Instead of treating GCG as a real substance, we only regard it as an effective description and convenient way to track down entropy evolution. The details of our model, including parameter space, are discussed in section (\ref{model}). We believe, however, that our result is quite general; see section (\ref{conclusion}).

\bigskip

\section{Generalized Chaplygin Gas\label{chap}}

\subsection{The Model}\label{model}
When applied to cosmology, the Chaplygin gas\cite{Lopez2011} models the change of cosmic content by regulating the equation of state of the background
fluid instead of the form of the potential. It was first suggested by Kamenshchik\cite{Kamenshchik2002} in an attempt to smoothly interpolate the de Sitter phase and the radiation dominant era without \textit{ad hoc} assumptions. Generalized Chaplygin gas models (GCG) were subsequently suggested\cite{Bento2002}\cite{Bertolami2004}\cite{Lopez2010}, some of which managed to unify dark energy and dark matter\cite{Bilic2002}. 
Instead of endowing it as a physical substance, we will only invoke it as an effective description.

The fluid density of GCG is given by\cite{Lopez2011}:
\begin{equation}\label{eqn:gcgdensity}
\rho=\Big(\frac{A}{a^{1+\beta}}+\frac{B}{a^{4(1+\gamma)}}\Big)^{\frac{1}{1+\gamma}},
\end{equation}
where $A$ and $B$ are positive constants. Note that in this expression we have implicitly taken a reference scale factor at some point during inflation to be $a_0=1$. One can also manifest this reference scale by rescaling $A\mapsto A a_0^{1+\beta}$, $B\mapsto B a_0^{4(1+\gamma)}$.

Consider the parameters $\beta=-1$ and $\gamma=-2$. Observe that when taking the limit $a\to 0$, one recovers from Eq.(\ref{eqn:gcgdensity}) the fluid density of de Sitter space characterizing the inflation era; when taking $a\to \infty$, the fluid density becomes that of a gas in the radiation-dominated era. Thus the two eras can be smoothly patched under this effective description. The constraints on the parameters $\beta$ and $\gamma$ to realize this scenario will be discussed in Section \ref{parameters}. 

From the conservation of energy one can deduce the pressure for the Chaplygin gas as
\begin{equation}
p= \frac{1}{3}\rho + \frac{1 + \beta -4(1 + \gamma)}{ 3(1 + \gamma)}
\Big(\rho-\frac{B}{ a^{4(1+\gamma)}}\rho^{-\gamma}\Big).
\end{equation}
After eliminating the scale factor $a$, one can obtain the equation of state for generalized Chaplygin gas.

One can also model GCG with an underlying minimal coupling scalar field $\phi$ by matching the energy momentum with the fluid described above.
Denoting the scalar potential by $V(\phi)$, we have:
\begin{equation}
\rho=\frac{1}{2}\dot{\phi}^2+V(\phi),\quad\quad\\
p=\frac{1}{2} \dot{\phi}^2-V(\phi).
\end{equation}
The classical expression for the scalar potential as well as the equation of motion can be immediately obtained by relating $a$ and $\phi$, as in \cite{Lopez2011}. For this purpose, we only need the limiting case for the early-time and the late-time approximations, respectively, with the partition roughly coincides with the crossing between the inflation and the radiation-dominated epochs. These are
\begin{itemize}
\item Early time [$a/a_0\ll\left(A/B\right)^{1/\delta}$]: \label{eqn:earlytimeapprox}
\begin{align}
&\phi(a)\sim \phi_0+\frac{1}{\kappa}\sqrt{\frac{1+\beta}{1+\gamma}}\ln\frac{a}{a_0},\label{eqn:earlyeom}\\
&V(\phi)\sim\left[1-\frac{1+\beta}{6(1+\gamma)}\right]A^{1/(1+\gamma)}a_0^{(\lambda/\kappa)^2} e^{-\lambda (\phi-\phi_0)},\label{eqn:earlypot}
\end{align}
where $\lambda\equiv \kappa\sqrt{(1+\beta)/(1+\gamma)}>0$ is the coupling constant, $\phi_0=\phi(a_0)$, and $\kappa^2 \equiv 8\pi G$.

\item Late time [$a/a_0\gg\left(A/B\right)^{1/\delta}$]: \label{eqn:latetimeapprox}
\begin{align}
&\phi\sim \phi_0'+ \frac{2}{\kappa}\ln\frac{a}{a_0},\label{eqn:lateeom}\\
&V(\phi)\sim\frac{1}{3}B^{1/(1+\gamma)}a_0^4 e^{-2\kappa(\phi-\phi_0')}.\label{eqn:latepot}
\end{align}
\end{itemize}

\subsection{Parameter Space of GCG}\label{parameters}
In this subsection we will briefly summarize the constraints on the generalized Chaplygin gas (\ref{eqn:gcgdensity})

First we require that the inflation takes place before radiation dominant era, which amounts to the condition:
\begin{equation}
\frac{B}{a^{4(1+\gamma)}}\ll\frac{A}{a^{1+\beta}}\label{restric1}
\end{equation}
as $a\to 0$. Since both $A$ and $B$ are positive, we have 
\begin{equation}\label{eqn:constraint1}
\delta\equiv 1+\beta-4(1+\gamma)>0.
\end{equation}
Two more conditions should be imposed
on $\beta$ and $\gamma$ so as to insure that our model does interpolate between
an early inflationary phase of the type of quintessence
and a subsequent radiation dominant phase. These are: (i) the
energy density must induce a period of inflation
and (ii) the inflation should not cause a
super-inflationary expansion; i.e. $0< \dot{H}$. That is, a
super-accelerating phase of the universe, where the
energy density grows as the universe expands, should not occur. 

By combining these two ansatz with the above inequality (\ref{eqn:constraint1}), we
arrive at the following additional constraints:
\begin{eqnarray}\label{eqn:constraints}
1+\beta<0,\\ \nonumber
1+\gamma<0,\\ \nonumber
1+\beta-2(1+\gamma)>0. 
\end{eqnarray}
In particular, the last constraint results from the requirement for an inflationary period $\ddot{a}>0$ as $a\to 0$, which directly leads to the constraint on the coupling constant of the GCG scalar field (\ref{eqn:earlypot}),
\begin{equation}\label{eqn:constraintlambda}
\lambda<\sqrt{2}\kappa.
\end{equation}
It is possible to have inflation at intermediate value of $a$ (see Appendix \ref{question}). However, in this paper we will only consider the case of inflation at the very beginning.

\subsection{Quantizing Generalized Chaplygin Gas}\label{qissuegcg}

Before entering into explicit calculations, we like to briefly comment on the problem one often encounters when implementing the quantization on the generalized Chaplygin gas model, such as the one we introduced in Section \ref{model}. When promoting a scalar field $\phi$ to a quantum field, one expects that the scalar potential would receive loop corrections. By inspecting on Eq. (\ref{eqn:earlypot}), one can quickly recognize that this model is nonrenormalizable by simply noticing the scalar coupling $\lambda$ has mass dimension $-1$. However, since the scalar field $\phi$ serves as the inflaton in the inflationary period, which decoheres entirely sometime after the onset of inflation, we do not expect the already decohered field to regain quantum fluctuations in the remaining period of inflation. We can therefore safely invoke the semiclassical approximation to the system beyond the point of decoherence, where one only has to consider the finite saddle-point contribution, and no regularization is required.
On the other hand, before the onset of decoherence of the inflaton, one does have to treat the scalar field $\phi$ that represents the Chaplygin gas fully quantum mechanically, thus some sort of regularization is still needed.

In our computation of the homogeneous part of the entropy, we invoke the minisuperspace approximation, which amounts to truncating large degrees of freedom and keeping only the homogeneous modes. Therefore the minisuperspace approximation serves as our regularization, based on which a finite result is obtained in Section \ref{qearlyentropy}
For the correct evaluation of the inhomogeneous part of the entropy at early times, one does have to impose a cutoff scale, below which our model is effective. We will discuss more on the issue of quantum corrections to the inhomogeneous entropy at early times at Appendix \ref{loopperturb}.

\section{Entanglement Entropy: homogeneous Part\label{WDW}}

As explained in section \ref{entropy}, the contribution to the cosmological entanglement entropy is divisible into two parts; in this section we will address the homogeneous contribution using the Wheeler-DeWitt (WDW) formalism and the minisuperspace treatment. The wavefunction of the universe depends on two modes: one from gravitation, the other from the scalar field of the Chaplygin gas with the spatial dependence suppressed, thus the name ``homogeneous''.

We will treat the gravitational degree of freedom using the flat FRW metric:
\begin{equation}\label{eqn:metric}
ds^2=-N^2dt^2+a(t)^2d\vec{x}^2.
\end{equation}
Throughout this paper we will follow the convention $\kappa^2\equiv 8\pi G=6$.

The action of our model is $S=S_{\mathrm{EH}}+S_{\mathrm{GCG}}$, where the Einstein-Hilbert action can be written as
\begin{equation}
S_{EH}=-\int\!d^3\vec{x} dt N\frac{1}{2}a^{-1}P_a^2,
\end{equation}
where $P_a=\dot{a}a/N$ is the canonical momentum conjugated with $a$. The spatial integral gives a volume $v_0$, which should not concern us and will be dropped hereafter.
The matter action is
\begin{equation}
S_{\mathrm{GCG}}=\int d^3\vec{x}dt Na^3\left( \frac{1}{2N}\dot{\phi}^2-V(\phi) \right),
\end{equation}
where $V(\phi)$ is the scalar potential of the Chaplygin gas.

Varying with the lapse function $N$, one can obtain the Hamiltonian equation $H=0$,
which can be quantized straightforwardly using the canonical prescription.
Hereafter we will stick to the gauge $N=1$.

\subsection{Semiclassical Calculation of Entanglement Entropy}\label{semiclentropy}

As mentioned in Section \ref{qissuegcg}, since the scalar field $\phi$ decoheres sometime during inflation, we can substitute $\phi$ by $a$ using the equation of motion. Therefore, the matter action is replaced essentially by the fluid density, and the Hamiltonian equation boils down to
\begin{equation}
H=-\frac{1}{2}a^{-1}P_a^2+a^3\rho(a)=0,
\end{equation}
which, after the canonical quantization $P_a\mapsto -i\frac{\partial}{\partial a}$, becomes
\begin{equation}
\hat{H}\Psi=\left[\widehat{\mathcal{O}}_{\mathrm{Grav}}+{\mathcal{O}}_{\mathrm{Matter}}\right]\Psi=0,
\end{equation}
where $\Psi=\Psi(a)$ is the wavefunction of the universe, $\hat{}$ denotes differential operators, and $\mathcal{O}_{\mathrm{Matter}}$ equals essentially to the generalized Chaplygin gas density (\ref{eqn:gcgdensity}). 
The equation above is the Wheeler-DeWitt equation with one dynamical degree of freedom $a$. For later convenience, we will factor out a $a^{-1}$ factor in both terms, making it resemble the Schr\"odinger equation for a unit mass particle.

Since we would like to compute the von Neumann entropy from the matter wavefunction, which can be extracted from the Born approximation \cite{Halliwell}
\begin{equation}\label{eqn:born}
\Psi\approx e^{iS_{\mathrm{EH}}}\psi,
\end{equation}
where $e^{iS_{\mathrm{EH}}}$ is the gravitational part, which, to the leading order, satisfies the Hamilton-Jacobi equation
\begin{equation}\nonumber
\left(\frac{dS_{EH}}{da}\right)^2-2a^4\rho(a)=0.
\end{equation}

The matter wavefunction $\psi$ satisfies the Schr\"odinger equation corresponding to the potential of the scalar field:
\begin{equation}\label{eqn:scalar_potential}
V(\phi(a))=V(a).
\end{equation}
To simplify the computation, we will invoke the early-time and the late-time approximations, i. e., Eq.(\ref{eqn:earlypot})and Eq.(\ref{eqn:latepot}), respectively. In particular, the potential has the asymptotic forms:
\begin{align}
&V(a)\sim V_0 \left(a/a_0\right)^{-(1+\beta)/(1+\gamma)},\quad \text{Early-time}\\
&V(a)\sim V_0' \left( a/a_0\right)^{-4},\qquad \qquad \quad \text{Late-time}
\end{align}
with $V_0$ and $V_0'$ as well as the criterion of division given in Section \ref{model}.
We observe that the potential is positive. Therefore the stationary ground state wavefunction decays exponentially.

Since this is a semiclassical treatment,we are only allowed to obtain the wave function through the WKB approximation:
\begin{equation}\label{eqn:wkb}
\psi(a) \sim Ce^{ -\int_{a_i}^a \sqrt{2V(x)}dx},
\end{equation}
where $a_i$ is a point in time before the semiclassical approximation breaks down. (cf. Section \ref{wkbbad})

From the above equation we can compute the von Neumann entropy with the density matrix $\bar{\rho}(a_1,a_2)=|\psi(a_1)|^2\delta(a_1-a_2)$, 
\begin{equation}\label{eqn:homentvon}
\Delta S(a) = -\int_{a_i}^a \! \bar{\rho} \ln\bar{\rho} \mathrm{d}\tilde{a}.
\end{equation}
A bar is used to distinguish it with the fluid density. As explained in section \ref{entrophori}, the integration over the range $\tilde{a}<a$ amounts to tracing out the inaccessible Hilbert space associated with the part of the universe outside of the particle horizon. Note that the real physical space has no direct relation to the fibration Hilbert spaces; furthermore, in the minisuperspace treatment the inaccessible universe though can be infinite in volume, has only one Hilbert space described by finite degrees of freedom, therefore the range of integration is compact.

\begin{figure}[h!]
\includegraphics[width=0.5\textwidth]{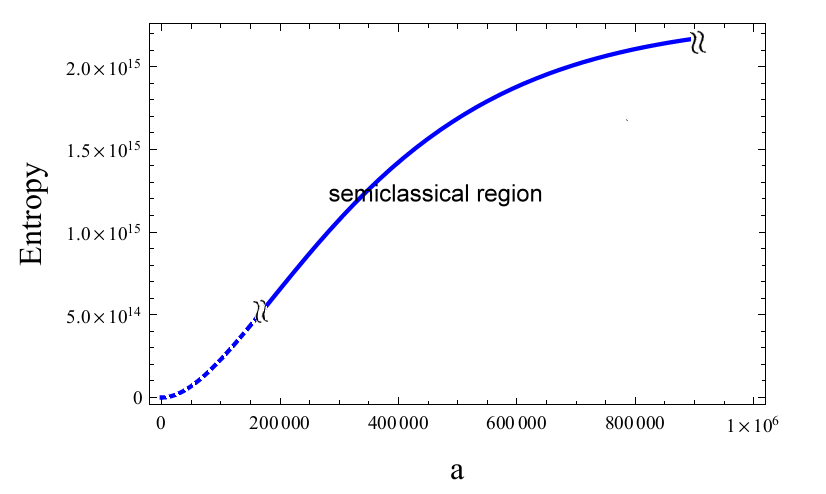}
\caption{The evolution of the homogeneous entanglement entropy under the early-time approximation, Eq.(\ref{eqn:earlyeom}) \& Eq.(\ref{eqn:earlypot}), contributed from the entanglement between the observable universe and that outside the horizon, obtained from the WKB wavefunction of Wheeler-DeWitt equation Eq.(\ref{eqn:wkb}).}\label{initialentropy}
\end{figure}

\begin{figure}[h!]
\includegraphics[width=0.5\textwidth]{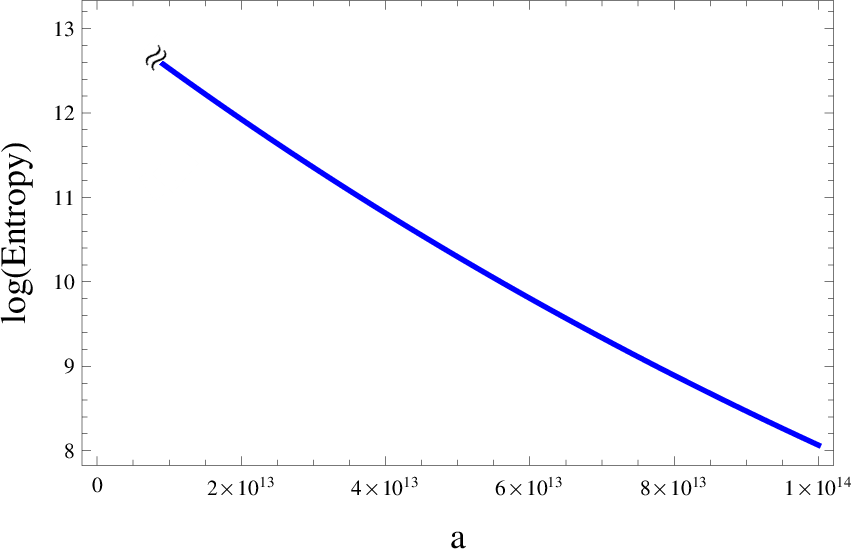}
\caption{The evolution of the homogeneous entanglement entropy under the late-time approximation within the radiation dominant era, Eq.(\ref{eqn:lateeom}) \& Eq.(\ref{eqn:latepot}), using von Neumann prescription Eq. (\ref{eqn:homentvon}). The tendancy indicates the entanglement entropy should start to decrease during transition from inflationary epoch.}\label{radiation}
\end{figure}

The numerical results are shown in Figure \ref{initialentropy} for early times and Figure \ref{radiation} for late times, where the late time is entirely in the radiation-dominated epoch.

From the figures one can observe that the evolution of the ``homogeneous entropy'' beyond the gravitational decoherence point is roughly increasing followed by decreasing starts from the transition to radiation epoch. This is expected since particle horizon only present in inflationary period, thus entanglement entropy is expected to decrease after some time in the transition.


\subsection{Comparison to the Area Law}\label{arealaw}

As mentioned in section \ref{entrophori}, the entanglement entropy of expanding universe was found in literature to be proportional to the area of the particle horizon. We will show that the homogeneous entanglement entropy computed in last section conforms with the area law.

The radius of the particle horizon $r_H$ is given in section \ref{entrophori} as
\begin{equation}\label{eqn:lightdistance2}
r_H[a]=\int_{t_0}^{\infty} \frac{(\mathrm{d}a/\dot{a})}{a}=\int_{a}^{\infty} \frac{\mathrm{d}\tilde{a}}{H(\tilde{a}) \tilde{a}^2}.
\end{equation}

\begin{figure}[H]
  \centering
    \includegraphics[width=0.45\textwidth]{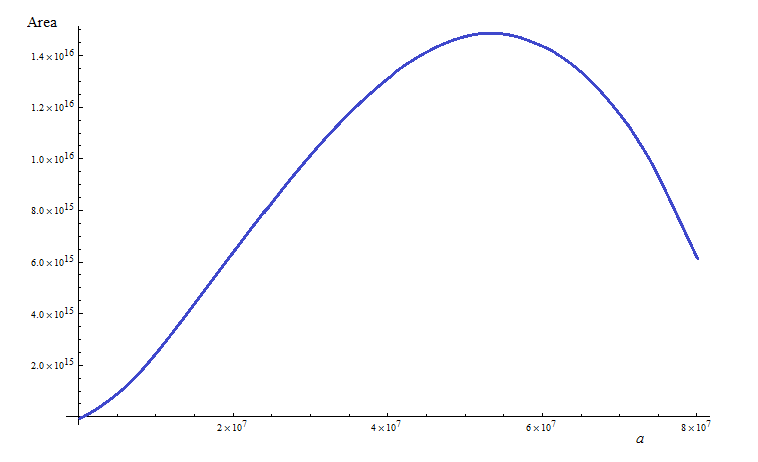}
  \caption{Evolution of horizon area before the radiation epoch}
    \label{fig:harea}
\end{figure}

Since particle horizon is well-defined only during the inflationary epoch, the upper limit is cut off at some time during the transition from inflation to radiation eras. Using Friedman equation and fluid density Eq.(\ref{eqn:gcgdensity}), we obtain the horizon boundary area $A_H= 4\pi \left(a r_H[a]\right)^2$ as a function of $a$ (see figure \ref{fig:harea}).

Comparing with the previous result of homogeneous entanglement entropy in figure \ref{initialentropy} and figure \ref{radiation}, we can see that the character of entropy evolution in the semi-classical regime agrees well with the area law: at first its growth accelerates, then it slows down, and finally it decays when entering the radiation epoch. We should emphasis that the notion of particle horizon is only well-defined in inflationary epoch much earlier than the transition, such as the case of early half of the evolution curve. However, in the entire semi-classical regime before radiation dominance, the conformation of our method with the area law is clearly established.

\subsection{Breakdown of Semiclassical Treatment}\label{wkbbad}
As has been pointed out, the semiclassical treatment of Chaplygin gas, as used in most literature, breaks down at the very beginning of the universe. Physically, we expect that at the onset of inflation the inflaton should execute large quantum mechanical fluctuations that seed the CMB spectrum. One should therefore treat the scalar field quantum mechanically before the decoherence of inflation occurs.

Consider the scalar potential after substituting the saddle point relation, $\phi=\phi(a)$, which at early times is of the form $V(a)\sim a^{-(1+\beta)/(1+\gamma)}$. The validity of the semiclassical (WKB) approximation requires the quantum potential to vary sufficiently slowly. To be more precise, it requires the WKB characteristic length
\begin{equation}
l_{\text{WKB}}\sim \frac{\hbar}{p_{cl}}=\frac{\hbar}{\sqrt{2V}}
\end{equation}
to change slowly over the distance, that is, 
\begin{equation}
\frac{dl_{\text{WKB}}}{da}\ll 1,
\end{equation}
where without loss of generality we have taken the mass to be 1.

When inserting the early time potential in Eq.(\ref{eqn:scalar_potential}), we find
\begin{equation}
\Big|{\frac{d\l_{\text{WKB}}}{d a}}\Big|  \sim  \Big|   \frac{d\left(V^{-1/2}\right)}{da} \Big|\sim a^{-\frac{1+\beta-2(1+\gamma)}{2(-1-\gamma)}}.
\end{equation}
However, with the constraints on parameters in Eq.(\ref{eqn:constraints}), it is easy to see that the right hand side goes to infinity when $a\to 0$, thus violating the validity condition of WKB semiclassical approximation.
A full quantum treatment of both $a$ and $\phi$ is therefore necessary for early-time universe, which we will carry out in the following Section \ref{earlyquantum}.

\subsection{Wavefunction at the Very Beginning of Universe}\label{earlyquantum}

We will implement a fully quantum mechanical treatment at early times via the full Wheeler-DeWitt equation:
\begin{equation}
\left[ \hat{\mathcal{O}}_{\text{Grav}}+\hat{\mathcal{O}}_{\text{Matter}}\right]\Psi(a,\phi)=0,
\end{equation}
where we have promoted $\phi$ to be an independent operator. Written explicitly, we have (in the convention $\kappa^2=6$)
\begin{align}
\left(\frac{1}{2}a\frac{\partial}{\partial a}
a\frac{\partial}{\partial a}-\frac{1}{2}\frac{\partial^2}
{\partial\phi^2}+a^6 V(\phi)\right)\Psi(a,\phi)=0,
\end{align}
or, with $\alpha=\ln\;a$, 
\begin{align}
\left(\frac{1}{2}\frac{\partial^2}{\partial\alpha^2}-
\frac{1}{2}\frac{\partial^2}{\partial\phi^2}+
e^{6\alpha}V\left(\phi\right)\right)\Psi(\alpha,\phi)=0.\label{eqn:wdw}
\end{align}

In order to recover the Chaplygin gas density at the saddle point, we adopt the scalar potential $V(\phi)=V_0 e^{\lambda \phi}$ from Eq.(6).
Thus the WDW equation is just the scalar field with an exponential potential in the FRW metric. This model is known to be exactly solvable using the following change of variables\cite{Phantom}:
\begin{align}
&x=\frac{\sqrt{2V_0}}{3}
\frac{e^{3\alpha-\lambda\phi/2}}{1-\left(\lambda/6\right)^2}\left[
  \cosh\theta(a,\phi)+\frac{\lambda}{6}\sinh\theta(a,\phi)\right],\\
&y=\frac{\sqrt{2V_0}}{3}
\frac{e^{3\alpha-\lambda\phi/2}}{1-\left(\lambda/6\right)^2}\left[
  \sinh\theta(a,\phi)+\frac{\lambda}{6}\cosh\theta(a,\phi)\right],
\end{align}
where $\theta(a,\phi)\equiv 3\phi-\lambda \alpha/2$.
The WDW equation Eq.(\ref{eqn:wdw}) becomes
\begin{equation}
\left[\frac{\partial^2}{\partial x^2}-\frac{\partial^2}{\partial y^2}+1\right]\Psi(x,y)=0,
\end{equation}
which can be easily solved to give
\begin{equation}
\Psi=\int\!\text{d}k \left[ C_1(k) e^{i\left(kx-\sqrt{k^2-1}y\right)} +C_2(k) e^{-i\left(kx-\sqrt{k^2-1}y\right)} \right].
\end{equation}

\begin{figure}[h!]
  \centering
    \includegraphics[width=0.5\textwidth]{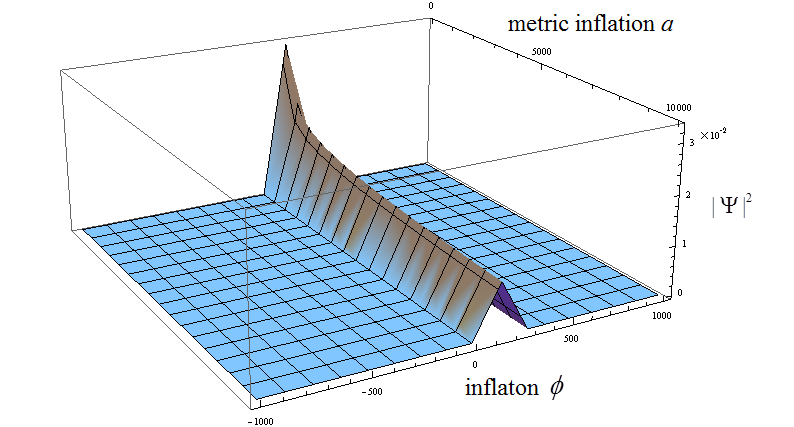}
   \caption{Distribution of the wavepacket solution Eq.(\ref{eqn:distrib})}   
    \label{fig:distribb}
\end{figure}

Take a Gaussian wave packet data with the peak at $k_0$ and width $\sigma$, we obtain a wave packet solution
\begin{align}\label{eqn:distrib}
|\Psi|^2=&\frac{2}{\sqrt{\pi}}\frac{(k_0^2-1)^{3/2}\sigma}{\sqrt{(k_0^2-1)^3+\sigma^4 y^2}}\nonumber\\
&\times\exp\left[ -(k_0^2-1)^2\sigma^2 \frac{\left(\sqrt{k_0^2-1}x-k_0 y\right)^2}{(k_0^2-1)^{3}+y^2\sigma^4} \right].
\end{align}
A typical profile of $|\Psi|^2$ is shown in Figure \ref{fig:distribb}. We can see that the distribution is peaked at the saddle point with a nonzero quantum-mechanical dispersion similar to the free field wavepacket in \cite{Kieferwp}.

The central value $k_0$ can be determined from the classical solution $\phi \sim \frac{\lambda}{6}\alpha$ (we take $\phi=0$ at $\alpha=0$) by identifying the location of the ``ridge'' where the distribution maximizes at any slice of a given $a$. The result is $k_0=-6/\sqrt{36-\lambda^2}$. 
Note that in our convention $\kappa=\sqrt{6}$, $\lambda<\sqrt{2}\kappa\approx 3.46$, which is below 6 and therefore $k_0$ is real and finite.

\subsection{Revised Calculation of Early-time Entanglement Entropy}\label{qearlyentropy}

\begin{figure}[h!]
  \centering
    \includegraphics[width=0.5\textwidth]{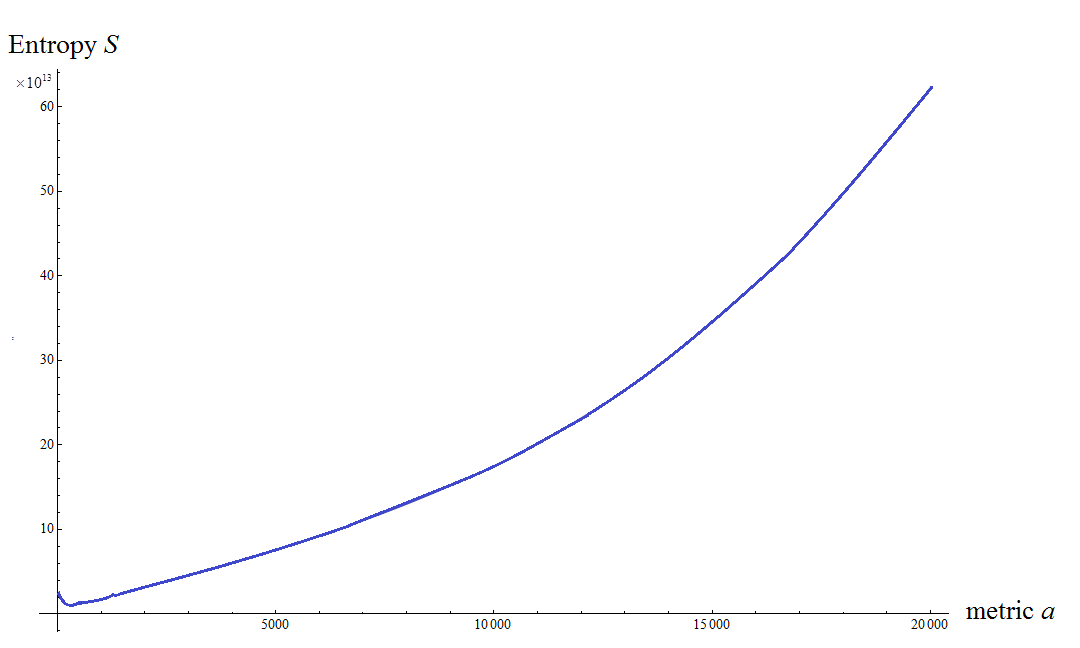}
   \caption{Early time homogeneous entanglement entropy evolution with quantum correction}   
   \label{fig:earlyentrop}
\end{figure}

With the wavefunction distribution, Eq.(\ref{eqn:distrib}), at hand, we can compute the von Neumann entropy $S=-\int\!\mathrm{d}a \mathrm{d}\phi |\Psi|^2 \ln\;|\Psi|^2$. The Hilbert space we are tracing over now has an additional degree of freedom $\phi$. Note that if our wavepacket has a well-defined peak with respect to $\phi$, then we will recover the equation of motion (\ref{eqn:earlyeom}) by the saddle point approximation upon the $\phi$ integration, and the entropy will be the same as in section \ref{semiclentropy}.
The resulting entropy profile is shown in figure \ref{fig:earlyentrop} (for $\lambda=3.3,\sigma=0.8$).

We see here that the general tendency of the entropy evolution is its accelerating growth, which matches well with the semi-classical result in Figure \ref{initialentropy}. This is expected since we use a well-defined wavepacket. 

In Figure \ref{fig:earlyentrop}, there is an early period where the numerical computation becomes less stable; however, within the range of tolerance one can observe the existence of an entropy minimum in the early period of inflation. In contrast, the horizon boundary area grows monotonically without this unusual decrease; the area law thus seems to be violated by quantum fluctuations described by the early universe wavefunction.

Here we provide two explanations for the appearance of this local entropy minimum. At the onset of inflation, the quantum fields are out of equilibrium. Thus thermodynamics are not expected to hold exactly and the entropy decreases. Another possibility is that the minisuperspace approximation fails to hold at early times. This possibility can be tested by adding back degrees of freedom. In our following discussion of inhomogeneous entropy we will pursue this issue further.


\section{Entanglement entropy from cosmological perturbations\label{pertSect}}
 
In this section we will address the part of entanglement entropy induced by the entanglement between different momentum modes, which we call the ``inhomogeneous'' entropy. This contribution can be calculated using the standard cosmological perturbation techniques \cite{DecohEntroI} \cite{DecohEntroII}. The calculation in this section largely follows \cite{Y}.

\subsection{Computation of Entropy from Cosmological Perturbation}\label{treeperturb}

Let us denote by $\nu$ the usual Mukhanov-Sasaki variable. The comoving curvature perturbation is then $\zeta(x, t)=\nu(x, t)/z(x, t)$, with 
$z(x, t)=a\sqrt{\epsilon}/4G$ and 
\begin{equation}
\epsilon=-\mathrm{d}\ln H/\mathrm{d}\ln a
\end{equation}
is one of the slow-roll parameters.
The conjugate momentum of $\zeta$ is denoted as $\pi=\partial_{\mu}\zeta$. The Gaussian random state in a single scalar field inflation is then characterized by the covariant matrix between two momentum modes $\textbf{k}$ and $-\textbf{k}$, which is related to the density matrix $\bar{\rho}$ by \cite{DecohEntroI}:
\begin{equation}\label{eqn:cov}
C=\mathrm{Tr}(\bar{\rho}{V V^{\dagger}})=\left( \begin{array}{cc}
P_\zeta&P_{\zeta\pi}\\
P_{\zeta\pi}&P_\pi,\\
\end{array} \right)
\end{equation}
where the trace is taken over the functional space, and 
\begin{equation}\nonumber
V=\left( \zeta_{\mathbf{k}},\pi_{-\mathbf{k}}\right),
\end{equation}
where $\pi=2a^3\epsilon\dot{\zeta}$ is the momentum conjugate with $\zeta$.

Each component in Eq.(\ref{eqn:cov}) can be identified (to tree level) as:
\begin{eqnarray}
P_\zeta(\mathbf{k},t)=|\zeta(\mathbf{k},t)|^2,\\
P_{\zeta,\pi}(\mathbf{k},t)=\frac{a^3 \epsilon}{4\pi G}\mathrm{Re}(\zeta(\mathbf{k},t) \partial_t\zeta(\mathbf{k},t)^\star),\\
P_{\pi}(\mathbf{k},t)=(\frac{a^3 \epsilon}{4\pi G})^2 |\partial_t\zeta(\mathbf{k},t)|^2.
\end{eqnarray}
We can then evaluate the entanglement entropy from cosmological perturbation by the equations:
\begin{eqnarray}\label{eqn:inhomentropy}
&S=2\sum_{\mathbf{k}}\left[(n_{\mathbf{k}}+1)\ln(n_{\mathbf{k}}+1)-n_{\mathbf{k}}\ln n_{\mathbf{k}})\right],\\
&(n_{\mathbf{k}}+\frac{1}{2})^2=\det C_{\mathbf{k}}.
\end{eqnarray}
The field $\nu(\mathbf{k},t)$ satisfies in Fourier space the equation
\begin{equation}
\ddot{\nu}+\frac{3\nu}{t}\dot{\nu}+\frac{k^2}{(F t)^{2\nu}}\nu=0,\label{perturbation}
\end{equation}
which governs the evolution of $\zeta(\mathbf{k},t)$. Here $F\equiv (1/(2\sqrt{3}\kappa))\lambda^2 A^{1/2(1+\gamma)}$ is a constant.

The basic steps to calculate the entropy from Eq.(\ref{eqn:inhomentropy}) are: (i) evolving the background metric by solving Einstein equation; (ii) solving the comoving curvature perturbation at time $t$ for different modes $\mathbf{k}$ according to Eq.(\ref{perturbation}); (iii) summing the entropy contributions from all modes which amounts to integrating Eq.(\ref{eqn:inhomentropy}) over the momentum space $\mathbf{k}$. In the calculation we employ the full tree-level scalar potential without early or late time approximation, given explicitly in \cite{Lopez2011}.

\begin{figure}[h]
\includegraphics[width=0.5\textwidth]{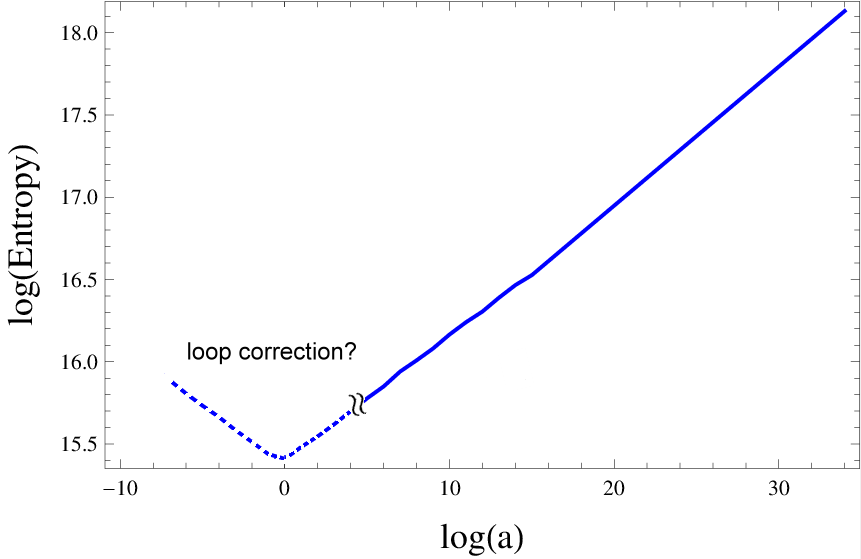}
\caption{Inhomogeneous entanglement entropy from tree-level cosmological perturbation}\label{fig:ppf}
\end{figure}

The numerical results are shown in Figure \ref{fig:ppf}. Observe that the tree-level contribution increases as power law in the radiation-dominant epoch started with $S_{\mathrm{inhom}}\sim 10^{16}$, which compensates almost entirely the exponential decrease of the homogeneous part (section \ref{semiclentropy}) of the entropy, which drops below $S_{\mathrm{hom}}\sim 10^{12}$ after $(a/a_0)\sim 10^{13}$. On the other hand, the homogeneous entropy during early times continues to increase, resulting in the monotonic rise of the net entropy in the semiclassical regime after the decoherence of the inflaton.

In the early stage of inflation there exists an entropy minimum, which confirms the result in Fig \ref{fig:earlyentrop} of Section \ref{qearlyentropy} that even after adding back some of the truncated degrees of freedom, there must still remain a persistent entropy minimum.
 We can therefore expect that the entropy minimum may actually persist to higher order quantum correction, rendering this turning point a real physical feature. We include some discussions about higher order correction in Appendix \ref{loopperturb}.

\section{Physical Origin of Entropy Decrease}\label{sec:explain}

In this section we present some general arguments on possible physical origin of the entropy decrease we found numerically in section \ref{qearlyentropy} and \ref{pertSect} with our model \ref{model}. 
The purpose here is to extend our argument without relying on any specific model, although our previous model is natural in the sense discussed in the beginning of section \ref{choice}, and in fact under the early and late time-scale approximations made in section \ref{semiclentropy} it is just a combination of very general power-law inflation.

The von Neumann entropy formula is
\begin{equation}\label{eqn:entropyintegral}
\Delta S(a):=-\int_{a_0}^a\int_{-\infty}^\infty \rho(a',\phi)\ln\rho(a',\phi)\mathrm{d}a'\mathrm{d}\phi.
\end{equation}
If the integrals are sufficiently well-behaved, i.e. uniformly convergent, then we can take the derivative with respect to parameter $a$, resulting in the change of entropy:
\begin{equation}\label{eqn:entropychnge}
\frac{d\Delta S(a)}{da}=\int_{-\infty}^\infty \rho(a,\phi)\ln[1/\rho(a,\phi)]\mathrm{d}\phi.
\end{equation}
Note that as long as $0<\rho<1$ and the integral converges, the above quantity stays positive, that is the entropy should increase monotonically. We comment here that such condition of monotonicity is a particular feature only in our situation, since the inaccessible environment depends on the metric expansion $a$ (which happens to coincide with cosmological time). In other quantum open systems, time will enter as external parameter in the density matrix $\rho(a,\phi;t)$, and in general one cannot obtain such a monotonic constraint of entropy from this simple argument.

There are two separate issues for this monotonicity:

\bigskip

\leftline{{\it 1. Convergence of entropy integral}}

If any of the entropy integrals Eq.(\ref{eqn:entropyintegral}) or Eq.(\ref{eqn:entropychnge}) has bad behavior such that some regularization should be imposed, or the integral in $\phi$ is oscillatory so uniform convergence is not guaranteed, then the above monotonicity argument fails. In the first case, the change of entropy will receive another ``boundary contribution'':
\begin{align}
&\frac{d\Delta S(a)}{da}=\int_{-\Lambda(a)}^{\Lambda(a)} \rho(a,\phi)\ln(1/\rho(a,\phi))\mathrm{d}\phi\\
 &\quad+\Lambda'(a)\left(\int_0^a \rho(a',\Lambda(a))\ln \rho(a',\Lambda(a))^{-1}\mathrm{d}a'+\right.\\
 &\quad\qquad\qquad \left.\int_0^a \rho(a',-\Lambda(a))\ln \rho(a',-\Lambda(a))^{-1}\mathrm{d}a' \right),
\end{align}
with the cutoff $\Lambda\to\infty$. However, this contribution can render the entropy variation negative if $\Lambda$ is a decreasing function of $a$. Physically, it means that as the cosmological time evolves the fluctuations of scalar field $\phi$ will be suppressed more and more, therefore it has less space for fluctuations. For the second case, the decrease of entropy is attributed not to any microscopic model, but to the stochastic nature of quantum physics.

\bigskip

\leftline{{\it 2. Unitarity}}
It has been a long debate whether the unitarity condition holds for the evolution of the universe. A well-known fact about QFT in curved background space-time is that a large number of particles can be created as the universe expands. Thus in our point of view the observable universe does not obey the unitarity condition. This is reminiscent of the Hawking radiation produced by the nontrivial black hole background. The Hawking particle so produced can be interpreted as an entanglement between the observable and unobservable regions in such background. Thus a natural proposition is that the universe {\it as a whole} should obey unitarity, but not separately in either observable or unobservable region. We therefore expect that the density matrix $\rho$ can exceed unity in either region, which is a natural consequence of particle production.

In the particle production (both matter and graviton) region, the change of entropy, Eq.(\ref{eqn:entropychnge}), is negative; whereas in the dissipative region, the change of entropy is positive. Putting this assessment into a concrete example, let us suppose a dissipative mechanism (say from coarse-graining) that results in an exponential decay of the total probability. Then the evolution of entropy can have the form displayed in Fig. 7.
\begin{figure}[h!]
  \centering
     \includegraphics[width=0.45\textwidth]{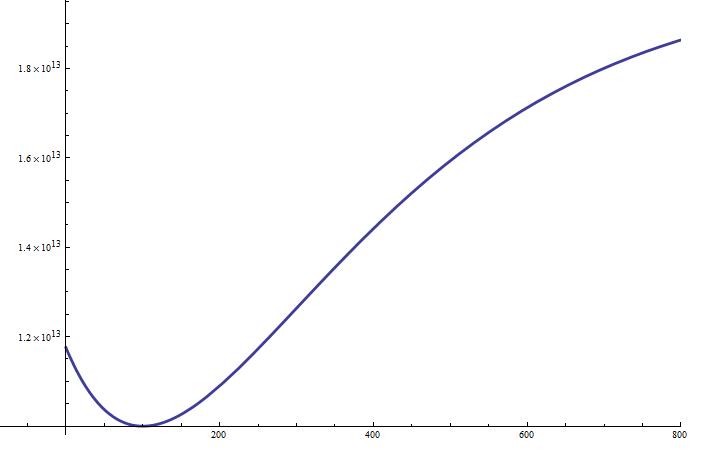}
      \caption{entropy evolution from a toy system}
\end{figure}

\bigskip

From the unitarity's point of view, the decrease of entropy boils down to the question of why our density matrix is so large at the very beginning? This is not too surprising since we expect quantum effects, which account for the particle production, to be prevailing near the Planck scale at the beginning of the universe. (Note that large density matrix does not imply the largeness of the total number of produced particles, but rather the {\it rate} of particle production.)
As the cosmological time $a$ elapses, the density matrix becomes smaller due to the slowdown of particle production rate. As discussed above, particle production is a process that drives the early universe out of equilibrium. Thus as time further evolves after such entropy decrease, the universe gradually recovers the equilibrium condition. It is only by then can one invoke second law of thermodynamics as the arrow of time.

\section{Conclusion}\label{conclusion}

In this paper we reconsidered inflation as a possible solution to the entropy problem on the basis neglected in the past literature that the probability of inflation, when taking its accompanied horizon into consideration, is not necessary small.
Indeed our computation suggests that the initial entanglement entropy should have a sizable value in order to compensate for its subsequent reduction at the beginning of inflation (cf. Fig. \ref{fig:ppf}), which indicates that the onset of inflation could be highly probable.

We explored this notion by investigating the evolution of entanglement entropy from the onset of inflation to the radiation era using a toy model, which we believe to be generic enough to capture the essential feature of entanglement entropy evolution because most calculations we made do not involve the model's artificial transition period between the two epochs, while at the same time it has the advantage of preserving unitarity instead of a discontinuous phase transition, which is disallowed in the quantum mechanical framework for the entire universe.

Our results of quantum entanglement entropy evolution are shown in Figure \ref{initialentropy} for homogeneous early evolution after decoherence of inflaton, Figure \ref{radiation} for homogeneous late-time evolution, Figure \ref{fig:earlyentrop} for homogeneous early evolution near the onset of inflation,
and Figure \ref{fig:ppf} for inhomogeneous evolution at the tree-level. The semi-classical evolution of entanglement entropy induced by the existence of particle horizon is shown to obey the area law, while near the onset of inflation this area law is found to be violated due to quantum fluctuations.

The net result for the period after decoherence of inflaton (in our case provided by the generalized Chaplygin gas at early time) is that the entanglement entropy increases monotonically, which results in a large amount of entropy generation and agrees with our experience with thermodynamic laws.

On the other hand, the situation before decoherence is more subtle. In particular, entropy minimum arisen from quantum corrections at early times appears in both the homogeneous part (Fig. \ref{fig:earlyentrop}) as well as the tree-level inhomogeneous part (Fig. \ref{fig:ppf}) previously observed in \cite{Y}. 
The persistence of the entropy minimum after putting back some of the degrees of freedom suggests that it is a real physical feature and we expect this entropy minimum 
not be completely erased by higher order corrections. 

In section \ref{sec:explain} we discussed possible physical origins that cause the entropy decrease and result in the subsequent entropy minimum: these include effective regularization on quantum fluctuation, particle production and dissipation. In particular, the latter two processes contribute to nonequilibrium effects during the early stage of accelerating expansion, rendering equilibrium physics such as the thermodynamic law of entropy temporarily inapplicable. By the same token, the equilibrium thermodynamic law gradually recovers as a result of a natural decay of such processes, thereby providing us the entropic arrow of time after the entropy reaches its minimum. 

We also like to comment on the possible connection of our investigation to other field-theoretic approaches such as in \cite{Srednicki}\cite{Susskind}. The decrease of entropy, which is a correction to the area law, can be compared to that in Eq.(\ref{eqn:ententropyscaling}) obtained in \cite{Susskind}. Moreover, if canonical quantum gravity can be regarded as a low energy limit of string theory, we expect that similar result may possibly be derived from supergravity or full string theory, although as remarked in Section \ref{qissuegcg}, a good UV description is lacking in our effective model.

Our results can be summarized as follows: after the onset of inflation the entropy will soon decrease, due to the dynamics of the expanding universe, to a minimum and rapidly increase thereafter. We therefore suggest that the special initial condition required for the thermodynamic arrow can be naturally provided by the inflation itself without the need to introduce any ad hoc assumptions.

\medskip

\bigskip

\begin{center}
\large {\bf Acknowledgment} 
\end{center}
We appreciate Yen Chin Ong, Je-An Gu and Chris Gauthier of the NTU LeCosPA Center for their useful suggestions. This research is supported by the National Science Council of Taiwan and the NTU LeCosPA Center.

\bigskip


\appendix

\section{Range of Parameters for Inflation}\label{question}
In this appendix we will show it is possible in our model to realize inflation starting from an expanding but non-inflationary stage, thereby the parameter space for inflation can be wider than considered in section \ref{parameters}, although in this paper we restrict to the case of inflation at the start.

For the inflation to start, the kinetic energy must be small in comparison with the potential energy \cite{McInnes2007}, which in our model amounts to the situation where the time derivative of the field
\begin{equation}
\frac{1}{2}\dot{\phi}^2= \frac{1}{2}\frac{d\phi}{da}\dot{a}^2 \sim H^2 \sim \frac{1}{2} Ca^{-(1+\beta)/(1+\gamma)}
\end{equation}
as well as its spatial derivative, $\frac{1}{2}k^2a^{-2}\phi^2$, be small in comparison with the potential
\begin{equation}
V\sim \left(1-\frac{1+\beta}{1+\gamma} \right) C a^{-(1+\beta)/(1+\gamma)}.
\end{equation}
The former requires 
\begin{equation}
1+\beta>3(1+\gamma),
\end{equation}
while the later will be the case after $a\gg \exp(1/r)$ with $r=2-(1+\beta)/(1+\gamma)$. Comparing to the requirement of early inflation at $a\to 0$ in Eq.(\ref{eqn:constraints}), we see that there is a finite nonzero range of parameters where the kinetic term is suppressed as the universe grows but not necessarily inflates, thereby paving the way for the system to become ``potential dominated'' and ready for the onset of inflation.

\section{Higher Order Quantum Corrections to Inhomogeneous Entropy}\label{loopperturb}

In this appendix we briefly discuss higher order corrections to the inhomogeneous entropy. 
Quantum corrections will only be important at early times. According to Eq.(\ref{eqn:earlypot}), the scalar field potential has the form
\begin{equation}
V=V_0 e^{-\lambda \phi},\label{eqn:earlypot2}
\end{equation}
where we have absorbed the constant $\phi$ into $V_0$.
Thus in comparison to the lowest order calculation, where only modes $k$ and $-k$ are entangled:
\begin{equation}
S=\int\!\mathrm{d}t\frac{\mathrm{d}^3k}{(2\pi)^3}\epsilon\frac{1}{2}a^3\left(\dot{\zeta}_k\dot{\zeta}_{-k}-a^{-2}k^2\zeta_k\zeta_{-k}\right),
\end{equation}
there is an additional entanglement due to the nonlinear interaction of Eq.(\ref{eqn:earlypot2}). This contribution can be partially accounted for by the noise kernel $N_k(t,t')$ and the diffusion kernel $D_k(t,t')$ \cite{DecohEntroII}:
\begin{align} 
  \frac{1}{2} & \frac{d\det(C)}{dt} = \left( \frac{a^3 \epsilon}{4\pi G} \right)^2 \cdot \nonumber\\
   &\left\{   \mathcal{P}_{\zeta\dot{\zeta}}(t)\,\int_{-\infty}^{t}\!dt' \, 
   \left[D_k G(t,t')-N_k G_{\mathrm{ret}}(t,t') \right] \right.\nonumber \\ 
 &\left.+ \mathcal{P}_{\zeta}(t)  
  \int_{-\infty}^{t}\!dt' \,
 \left[  N_k \partial_t G_{\mathrm{ret}}(t,t')-D_k\partial_t G(t,t') \right]\right\},\label{EvoldetC} 
\end{align}
with $N_k$ and $D_k$ obtained through the standard diagrammatic rules, where higher-order loop effect enters.

However, upon expansion of the potential Eq.(\ref{eqn:earlypot2}) one finds infinitely many types of vertexes, where each type will have its own coupling due to counter terms:
\begin{equation}
V_{\mathrm{quantum}}(\phi)= c_0+c_1\phi+c_2\phi^2+\cdots.
\end{equation}
This immediately leads to problems when performing perturbative loop calculations. Therefore numerical lattice computation seems desirable to obtain higher order corrections to early time inhomogeneous entropy.

%
%
%

\end{document}